# Prediction of butt rot volume in Norway spruce forest stands using harvester, remotely sensed and environmental data

Janne Räty[1*], Johannes Breidenbach[1*], Marius Hauglin[1], and Rasmus Astrup[1]

[1]Norwegian Institute of Bioeconomy Research (NIBIO), Høgskoleveien 8, 1433 Ås, Norway

[*]Correspondence: [first_name].[last_name]@nibio.no

## Abstract

Butt rot (BR) damages associated with Norway spruce (*Picea abies [L.] Karst.*) account for considerable economic losses in timber production across the northern hemisphere. While information on BR damages is critical for optimal decision-making in forest management, the maps of BR damages are typically lacking in forest information systems. We predicted timber volume damaged by BR at the stand-level in Norway using harvester information of 186,026 stems (clear-cuts), remotely sensed, and environmental data (e.g. climate and terrain characteristics). We utilized random forest (RF) models with two sets of predictor variables: (1) predictor variables available after harvest (theoretical case) and (2) predictor variables available prior to harvest (mapping case). We found that forest attributes characterizing the maturity of forest, such as remote sensing-based height, harvested timber volume and quadratic mean diameter at breast height, were among the most important predictor variables. Remotely sensed predictor variables obtained from airborne laser scanning data and Sentinel-2 imagery were more important than the environmental variables. The theoretical case with a leave-stand-out cross-validation achieved an RMSE of 11.4 $m^3 \cdot ha^{-1}$ (pseudo R²: 0.66) whereas the mapping case resulted in a pseudo R² of 0.60. When the spatially distinct k-means clusters of harvested forest stands were used as units in the cross-validation, the RMSE value and pseudo R² associated with the mapping case were 15.64 $m^3 \cdot ha^{-1}$ and 0.37, respectively. This indicates that the knowledge about the BR status of spatially close stands is of high importance for obtaining satisfactory error rates in the mapping of BR damages.

# 1 Introduction

Butt rot (BR) damages associated with coniferous forests account for considerable economic losses in the forestry sector of the northern hemisphere. BR damages are especially destructive in forests dominated by Norway spruce (*Picea abies* [L.] Karst). For example, it has been observed that 25 % of Norway spruce stems had BR damages in final fellings according to a Norwegian nation-wide stump survey (Huse et al., 1994). The most destructive fungus genus in coniferous forests is *Heterobasidion spp.* that comprises several species with varying host preferences and *Heterobasidion parviporum* especially prefers Norway spruce as a host tree species. In total, *Heterobasidion spp*. causes an annual economic loss of approximately 800 million euros in Europe alone (Hodges, 1999).

BR infections can spread from an infected tree to a healthy tree via airborne fungus spores that occupy a new contamination surface, such as a fresh stump. The BR infection can also enter to a new host tree via below-ground root connections (Aosaar et al., 2020; Stenlid, 1987). The spread of BR infection is also dependent on the characteristics of forest stands. It has been found that the risk of infection is lower in stands with a mixture of Norway spruce and Scots pine (*Pinus sylvestris* [L.]) than in pure Norway spruce forest (Möykkynen and Pukkala, 2010). Tree attributes, such as diameter at breast height (DBH) and age, are also linked to the risk of the BR damages (Hylen and Granhus, 2018; Mattila and Nuutinen, 2007). Hylen and Granhus (2018) found that the risk of BR damage in Norway spruce increases with DBH up to a DBH of 30 cm. They also found that the risk of BR damage increases in terms of age, but the probability of damage is relatively stable for trees older than 80 years. BR damages have frequently been found on calcareous, limestone-rich and fertile soil types, and it has been observed that a thick peat layer prevents the risk of the RB damages to some extent (Müller et al. 2018).

Cut-to-length harvesters collect tree-level data during harvest operations. Harvester datasets have been used joined with remotely sensed material and other auxiliary datasets to produce or validate forest resource information. Previous studies have used harvester data, for example, for the modeling of diameter distributions (Maltamo et al., 2019; Söderberg et al., 2021), the prediction of sawlog volume (Peuhkurinen et al., 2008) or other forest attributes (Söderberg, 2015; Hauglin et al., 2018) and the validation of forest attribute maps (Vähä-Konka et al., 2020). A major restriction related to the operational application of harvester data is the lack of standardized technology for the positioning of the trees (Hauglin et al., 2017; Kemmerer and Labelle, 2020), which is essential if harvester data are a surrogate for conventional field measurements.

The quality characteristics associated with standing timber resources are laborious to assess in the field. It is possible to detect quality attributes such as branchiness or crookedness (Karjalainen et al., 2019), but the accurate detection of BR damages by a visual inspection of standing trees is practically impossible. Therefore, harvester data can be important sources of quality characteristics associated with timber resources. Harvester data provide information on timber assortments at the level of logs cut from individual trees. Typically, the commercial timber assortments are sawlog, pulpwood, energy wood. Timber damaged by BR is typically allocated to the pulpwood or energy wood assortments, which results in economic losses especially in mature forest stands. In order to separate healthy and infected logs, visual inspections of cross-cutting surfaces during harvest operation must be carried out.

Currently, forest owners cannot accurately evaluate economic loss caused by BR damages in Norway. Thus, our objective was to map timber volume with BR damages (henceforth BR volume) in spruce-dominated forests observed from harvesters using remotely sensed and environmental variables. The remotely sensed variables comprised airborne laser scanning (ALS) and Sentinel-2 satellite imagery whereas the environmental variables consisted of climate and terrain variables, and site-specific characteristics indicating, for example, growing conditions, and geographical position. To the best of our knowledge, the application of harvester data in the mapping of BR damages jointly with remotely sensed variables, has not been studied so far.

## 2 Material and Methods

### 2.1 Study area

The study area is located between the latitudes of 59° and 65° in Norway. The large latitudinal range, changes in the distance to the coastline, and elevation shifts affect the growing conditions across the study area (Figure 1). The high-altitude (above sea-level) mountain forests were not of interest, since the focus was on the operationally accessible forests that are under commercial timber production. The mean altitude associated with the forests of interest was 300 m whereas the maximum altitude was 900 m. Norway spruce and Scots pine are the most common tree species in the area of interest. Broadleaved species, mostly birch (*Betula spp*. [L.]), are typically growing as mixtures among the coniferous species.

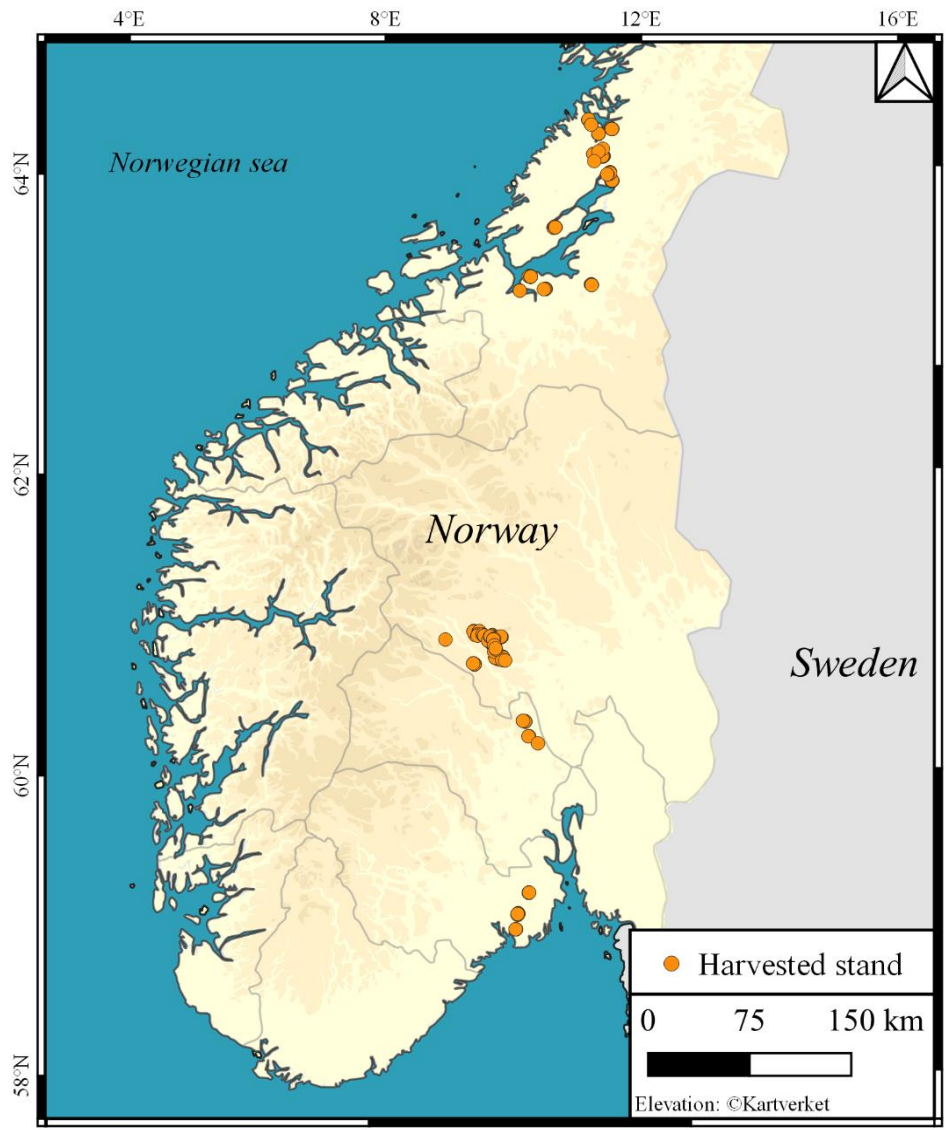

**Figure 1.** Study area and the locations of the harvested forest stands.

## 2.2 Harvester data

The harvester data consisted of 323,724 trees and were collected in 2020 and 2021 using five different harvesters. The harvester data only included trees from clear-cut stands. Few retention trees were usually left in the clear-cut stands, but they were not registered.

The harvester data comprise information on each harvested tree. The trees were bucked during the harvest operation into the commercial timber product categories sawlog, pulpwood and energy wood. The harvester sensors recorded diameter measurements along the stem including diameter at breast height (DBH), product lengths and product volumes (Nordström and Hemmingsson, 2018).

Operators of the harvesters manually recorded tree species and BR damages at each cross-cutting for each Norway spruce stem. The products damaged by BR were categorized into the products BR pulpwood, BR energy wood, and BR cut-off. The harvester data were stored in the Standard for Forest machine Data and Communication (StandForD) format (Arlinger et al., 2012).

BR volume was calculated for each Norway spruce stem. BR volume was calculated based on the damaged stem product or products, and infected stems usually comprise both damaged and healthy timber products.

## 2.3 Remotely sensed and environmental data

ALS data covering the study area were collected in several flight campaigns between 2010 and 2018. The flight parameters were not identical among the data acquisitions and the resulting mean point densities varied between 2–5 points per square meter among the ALS campaigns. A digital terrain model (DTM, $1 \times 1$ m) was created using the last returns of the ALS data (Kartverket, 2019). The DTM was subtracted from the orthometric height measurements of the ALS data to normalize them. The height-normalized ALS data were overlaid on the $16 \times 16$ m grid cells and the following ALS features were calculated based on first-of-many and only echoes for the cells of the SR16 map (Hauglin et al., 2021): mean, variance, proportion of echoes above 2 m, and percentiles ($25^{th}$, $50^{th}$, $75^{th}$, $90^{th}$, and $95^{th}$).

A mosaic of Sentinel-2 satellite imagery (Level 2A product) was obtained from the SentinelHub (Kirches, 2018). The mosaic was based on the images acquired in 2018. The medoid method (Flood, 2013) was used to obtain the most representative ground surface pixel value within each month, thus avoiding clouds and haze. After preliminary modeling attempts, we only used the B8 (near-infrared) band which has a spatial resolution of 10 m. Finally, the mosaic was resampled using bilinear interpolation to 16×16 m grid cells.

We collected environmental variables from several existing maps. Altitude ($AL_{CLI}$), temperature sum ($TS_{CLI}$), precipitation ($PS_{CLI}$), a terrain variable describing slope ($SL_{TER}$) and distance from the coastline ($DC_{CLI}$) were collected from existing nation-level maps ($16 \times 16$ m cell size) created for research purposes. The temperature sum was summed from mean of monthly mean temperatures (mean temperature threshold $> 5°$ Celsius) in the time period of 1989–2018. The precipitation was calculated as a sum of monthly precipitation of the months with mean temperature $> 5°$ Celsius in the same time period than the temperature sum. The temperature and precipitation data were downscaled from a 1×1 km cell size and the temperatures observations were also calibrated using elevation (Skaugen et al., 2002).

We collected soil characteristics ($ST_{AR5}$), and forest type ($FT_{AR5}$) classification from the Norwegian national land resource map AR5 map (Ahlstrøm et al., 2019). The $ST_{AR5}$ layer separates mineral soils from the organic soils, and the $FT_{AR5}$ layer describes the composition of stands as coniferous, deciduous, or mixed. The AR5 map is originally in a vector format and was converted to raster files that align with the 16×16 m grid cells. We also used the SR16 site index ($BON_{SR16}$) map (Astrup et al., 2019).

We also collected characteristics from the soil map (SOIL) provided by Geological Survey of Norway (2020). The SOIL map describes the soil type more specifically than the $ST_{AR5}$ map and has among others the categories the moraine and old sea-floor soil types that indicate high soil pH. The SOIL map is originally in a vector format and was converted to raster files that align with the 16×16 m grid cells.

## 2.4 Data preparation, modeling and validation

### 2.4.1 Study workflow

The methodology consisted of five steps: (1) post-processing of harvester data (2) the delineation of harvested stands, (3) the calculation of BR volume and predictor variables for the harvested stands, (4) the modeling of stand-level BR volume and (5) mapping and model validation using leave-stand-out and leave-cluster-out cross validation. The workflow is visualized in Figure 2 and the steps are explained in detail in the next sections.

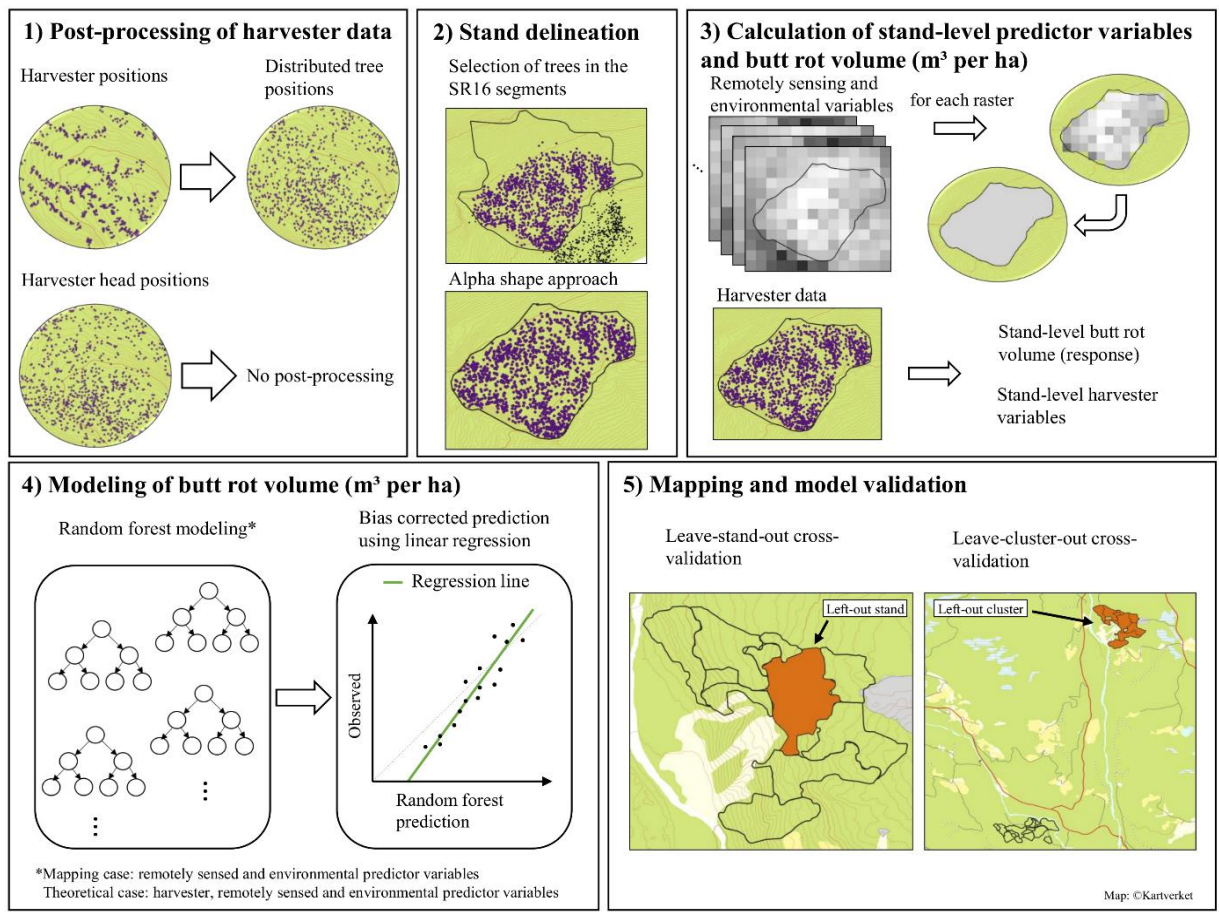

**Figure 2.** Study workflow.

### 2.4.2 Post-processing of tree locations

The harvesters were equipped with a global navigation satellite system (GNSS) receiver that registered the machine's location during harvesting operation. Three harvesters were equipped with a positioning system that determined the XY location of a harvester head which resulted in accurate tree positions (52 % of the trees). The XY locations of the harvested trees that were not positioned based on the harvester head followed stripe patterns in the harvested stands (Figure 2, step 1). In order to better distribute the tree locations for the delineation of harvested stands, we added a uniformly distributed random value of ±8 m to the XY coordinates of the machine to simulate the position of the harvester head. A preliminary analysis showed that the post-processing step of machine-based tree locations improved the delineation of stands from the harvester data.

### 2.4.3 Delineation of harvested stands

The SR16 map includes stand-like segments in forests and we used the SR16 segments as the base information for stand delineation (Figure 2, step 2). The SR16 segmentation is based on canopy height, as well as site index, and tree species composition predicted using remotely sensed and national forest inventory data (Astrup et al., 2019).

The harvested areas did not always match with the SR16 segments (Figure 2, step 2) and therefore, the XY locations of the harvested trees were utilized to crop of the SR16 segments. The harvested trees were attributed to SR16 segments using their XY locations. The XY locations of trees were used to create two-dimensional alpha shapes (n = 667) using the *alphahull* package (Pateiro-Lopez et al., 2019) in the R environment (R Core Team 2020). A buffer of two meters was added on the boundary of each alpha shape to account for the distance between crown edge and stem position. Finally, the SR16 segments were cropped by overlaying the alpha shapes to create segments of the harvested area. Henceforth, we refer to these cropped SR16 segments as *harvested stands*. The alpha parameter associated with the alpha shape approach was set to 25. Harvested stands with an area less than 0.3 hectares, with less than 30 harvested trees, and mixed stands with a volume proportion of Norway spruce less than 50 % were removed. Altogether, 256 spruce-dominated harvested stands with a total area 326 ha comprising 186,026 harvested stems were available for modeling BR rot volume. Statistics associated with the harvested stands are in Table 1.

**Table 1.** Statistics associated with the harvested stands.

|  | Mean | Standard deviation | Minimum | Maximum |
|---|---|---|---|---|
| Response variable: harvested volume damaged by butt rot ($m^3 \cdot ha^{-1}$) | 23.9 | 19.7 | 0.0 | 134.7 |
| Proportion of harvested volume damaged by butt rot (%) | 11.4 | 7.2 | 0.0 | 37.4 |
| Harvested volume ($m^3 \cdot ha^{-1}$) | 216.3 | 114.7 | 40.9 | 703.0 |
| Quadratic mean diameter (cm) | 22.0 | 3.3 | 14.2 | 32.3 |
| Number of harvested stems ($ha^{-1}$) | 743 | 260 | 203 | 1709 |
| Number of harvested stems per stand | 727 | 617 | 108 | 4471 |
| Harvested volume of Norway spruce (%) | 90.0 | 11.6 | 52.8 | 100.0 |
| Stand area (ha) | 1.0 | 0.8 | 0.3 | 5.1 |

*2.4.4 Calculation of butt rot volume and predictor variables for harvested stands*

Rasters of remotely sensed and environmental variables with $16 \times 16$ m grid cells were linked to the harvested stands by aggregating the grid cells with centers inside the boundaries of a harvested stand (Figure 2, step 3). Continuous variables were aggregated as means whereas the categorical variables were aggregated as mode values.

Predictor variables calculated from ALS and Sentinel-2 rasters will be referred to as *remotely sensed variables*. The variables calculated from the existing maps (e.g., SR16, AR5, SOIL) will be referred to as *environmental variables*. A summary of variables is presented in Table 2.

The stand-level response variable (BR volume) was calculated using the stem product information associated with the harvested trees. We also computed the following predictor variables from the harvester data: quadratic mean diameter of Norway spruce stems ($QMD_{HRV}$), harvested timber volume per hectare ($V_{HRV}$), the proportion of harvested Norway spruce volume, and the width of the DBH distribution ($DR_{HRV}$). The width of harvested DBH distribution was determined as the difference between the 90$^{th}$ and 10$^{th}$ percentiles of the DBH distribution. We henceforth refer to the predictor variables computed from the harvester data as *harvester variables*.

The harvester, remotely sensed and environmental variables were further categorized into two categories which were used to train two separate models: i) all predictor variables and ii) predictor variables available prior to harvest (i.e. harvester variables excluded). The former refers to the *theoretical case* with the availability of observed information on forest attributes because the harvester variables are available only after harvest. The latter category refers to the *mapping case* because these variables are available from the forest resource map SR16.

**Table 2.** Predictor variables calculated for the harvested stands.

| Available after harvest | | Available prior to harvest | | | |
|---|---|---|---|---|---|
| Harvester variables | | Remotely sensed variables | | Environmental variables | |
| Variable(s) | Description | Variable(s) | Description | Variable(s) | Description |
| $V_{HRV}$, $N_{HRV}$ | Harvested timber volume ($V_{HRV}$, m³·ha⁻¹) and the number of harvested stems ($N_{HRV}$, ha⁻¹) | $Hmean_{ALS}$, $Hvar_{ALS}$ | Mean, and variance associated with the height measurements of ALS data | $AL_{CLI}$, $SL_{TER}$, $TS_{CLI}$, $PS_{CLI}$, $DC_{CLI}$ | Altitude above sea level, slope, temperature sum ($TS_{CLI}$), precipitation ($PC_{CLI}$), and distance to coast ($DC_{CLI}$) |
| $QMD_{HRV}$ | Quadratic mean DBH of harvested Norway spruce stems (cm) | $HP_{ALS}$ | Percentiles associated with the distributions of height measurements of ALS data. P = {25, 95 %} | $BON_{SR16}$ | Site index extracted from the Norwegian forest resource map SR16 |
| $DR_{HRV}$ | Difference between the 90th and 10th percentiles of the DBH distribution. (cm) | $D2_{ALS}$ | Proportion of ALS height measurements (first echoes) above a threshold of 2 meters | $FT_{AR5}$, $ST_{AR5}$, SOIL | Forest type ($FT_{AR5}$) and soil type ($ST_{AR5}$) extracted from the Norwegian land resource map (AR5). The SOIL variable describes geological soil characteristics and was extracted from the map provided by the Geological Survey of Norway |
| $SPP_{HRV}$ | Proportion of harvested timber volume of Norway spruce (%) | $NIR_{S2}$, | Optical image variables extracted from the Sentinel-2 image mosaic. The following band was used: Near-infrared (NIR, band 8) | X, Y | X and Y coordinates of the centroids of the forest stands |

Note: HRV – harvester variable, ALS – airborne laser scanning, S2 – Sentinel-2, CLI – climate variable, TER – terrain variable, SR16 – Norwegian forest resource map, AR5 – Norwegian national land resource map

*2.4.5 Modeling butt rot volume*

We used Random Forest (RF) (Breiman, 2001) to model and subsequently map BR volumes at the stand-level. RF is a widely used non-parametric and non-linear approach which is based on classification and regression trees (CART). We used the RF implementation of the *randomForest* package (Liaw and Wiener, 2002) in the R environment. RF is controlled by three hyperparameters which determine the number of decision trees (*ntree*), the number of predictor variables selected in each node splitting (*mtry*) and the depth of a tree (*nodesize*). The hyperparameter values were fixed at their defaults of 500 and 5 for *ntree* and *nodesize*, respectively. The hyperparameter *mtry* is dynamically determined as the number of predictor variables divided by three brought up to a round integer. Preliminary analysis showed that changes in the hyperparameter settings only marginally affected the results.

It has been observed that RF regressions tend to overestimate small observations and underestimate large observations (Zhang and Lu, 2012). We reduced this prediction bias using a simple linear regression approach which relates observed values with RF predictions (Song, 2015). That means, our final predictions are based on an ordinary least squares regression model with the observed values as the response and the RF predictions as the only predictor variable (Figure 2, step 4).

*2.4.6 Validation and performance assessment*

We applied two different validation strategies in the performance assessment (Figure 2, step 5). In order to study the importance of close-by training data for the predictive performance, we used k-means clustering to create geographically independent groups of harvested stands. The k-means clustering was carried out the *stats* R-package (R Core Team, 2020). Only clusters with five or more harvested stands were allowed which resulted in 23 clusters. A distance to the center of the nearest cluster was on average 10 km, at minimum 0.5 km, and at maximum 57 km. There were on average 11 harvested stands per cluster. The resulting clusters were used to carry out a leave-cluster-out cross validation (ClusterCV). We also carried out a leave-stand-out cross validation (StandCV), which allows the inclusion of the geographically neighboring stands in the training data of the RF model.

We evaluated the predictive performance of the models using a pseudo-coefficient of determination ($R^2$):

$$Pseudo\ R^2 = 1 - MSE / \frac{\sum(y_i - \bar{y})^2}{n-1} \quad (1)$$

where $MSE = \frac{\sum_{i=1}^{n}(y_i - \hat{y}_i)^2}{n}$ is the mean squared error, and $y_i$ and $\hat{y}_i$ are observed and predicted BR volumes in stand $i$, $n$ refers to the number of harvested stands, and $\bar{y}$ is the mean of observed BR volume over all stands. For simplicity, we will refer to the pseudo $R^2$ value as $R^2$.

The error associated with predicted BR volume prediction was evaluated using the root-mean-square error (RMSE, Eq. 1) and mean deviance (MD, Eq. 2). The relative error is the absolute error divided by the observed mean of the response multiplied by 100.

$$RMSE = \sqrt{MSE} \quad (2)$$

$$MD = \frac{\sum_{i=1}^{n}(y_i - \hat{y}_i)}{n} \quad (3)$$

# 3 Results

## 3.1 Prediction of butt rot volume

Two cross validation strategies, namely StandCV and ClusterCV, were employed. The latter was used to evaluate the importance of close-by reference observations on the models' predictive performances.

The StandCV strategy resulted in smaller error rates than the ClusterCV strategy. With both cross-validation strategies, the exclusion of harvester variables increased the error rates associated with the predicted butt rot volumes. The exclusion of the harvester variables increased the RMSE values by 8.7 % and 8.9 %, for ClusterCV and StandCV, respectively. The magnitude of MD was moderate in all cases. Table 3 shows the RMSE and MD values associated with cross-validated predictions of BR volume. The predicted versus observed values given predictor variable and the cross-validation strategies are shown in Figure 3.

An example of BR mapping using the RF model and the SR16 segments is shown in Figure 4. The models presented in this study are not applicable for young forest stands. The "not applicable" stands shown in Figure 4 were filtered out by comparing the attributes associated with our training data and the attributes provided in the SR16 map (Not applicable: 95$^{th}$ percentile of ALS height distribution < 12 m and spruce volume proportion < 50 %).

**Table 3.** Root-mean-square errors (RMSE) and mean deviances (MD) associated with predicted volume damaged by butt rot using different sets of predictor variables and two different cross-validation (CV) strategies. The harvested stands were used as modeling units. ClusterCV – Leave-cluster-out CV, StandCV – Leave-stand-out CV

| CV strategy | Predictor variables | RMSE (m$^3$·ha$^{-1}$) | MD (m$^3$·ha$^{-1}$) | RMSE (%) | MD (%) | Pseudo-R² |
|---|---|---|---|---|---|---|
| ClusterCV | All | 14.38 | -0.64 | 60.11 | -2.69 | 0.47 |
|  | Prior to harvest | 15.64 | 0.07 | 65.36 | 0.30 | 0.37 |
| StandCV | All | 11.42 | -0.03 | 47.73 | -0.12 | 0.66 |
|  | Prior to harvest | 12.44 | 0.05 | 52.00 | 0.22 | 0.60 |

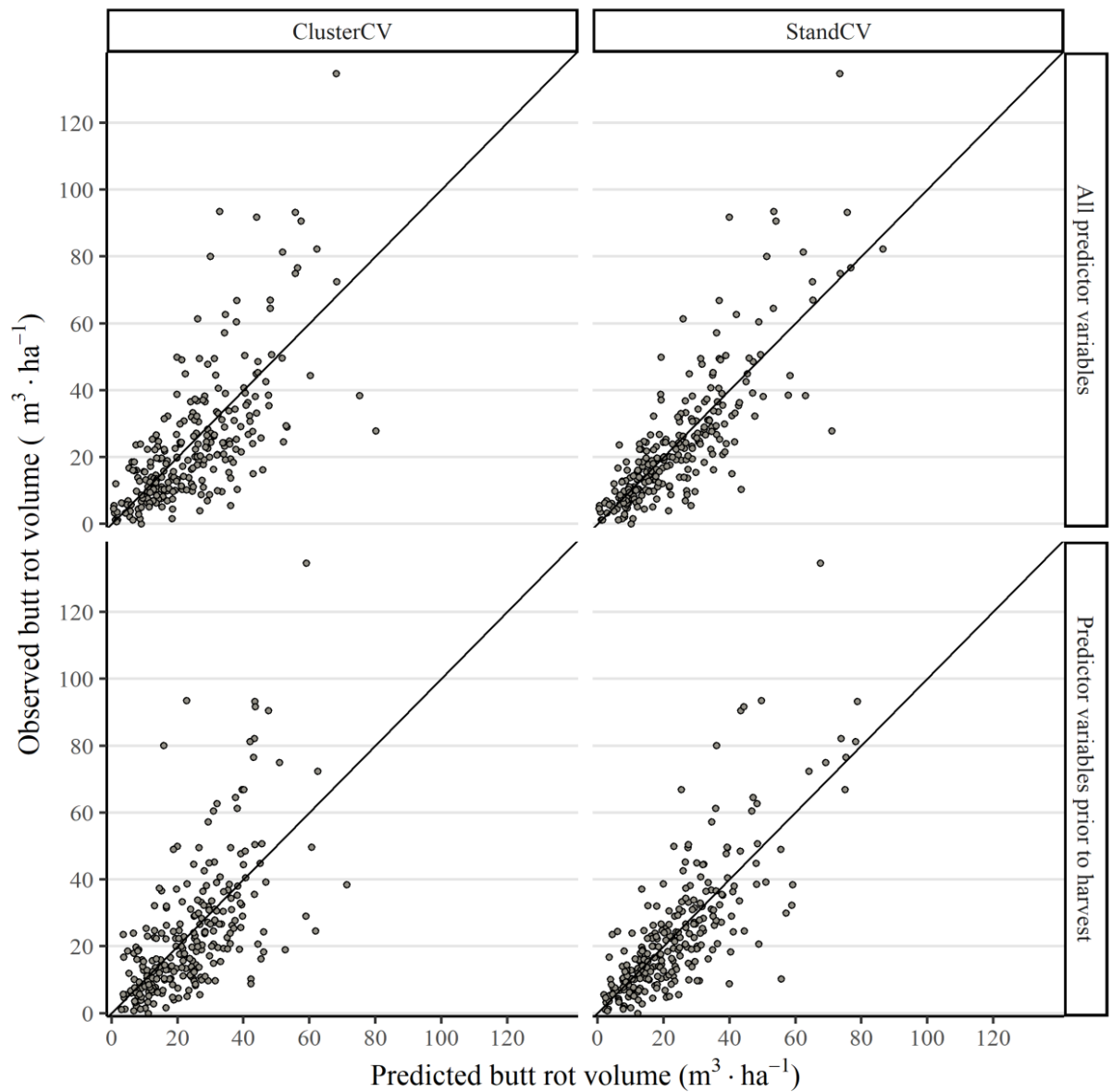

**Figure 3.** Observed versus predicted butt rot volume using leave-cluster-out (ClusterCV) and leave-stand-out cross-validation (StandCV) strategies. The top row: all available predictor variables; bottom row: predictor variables available prior to harvest.

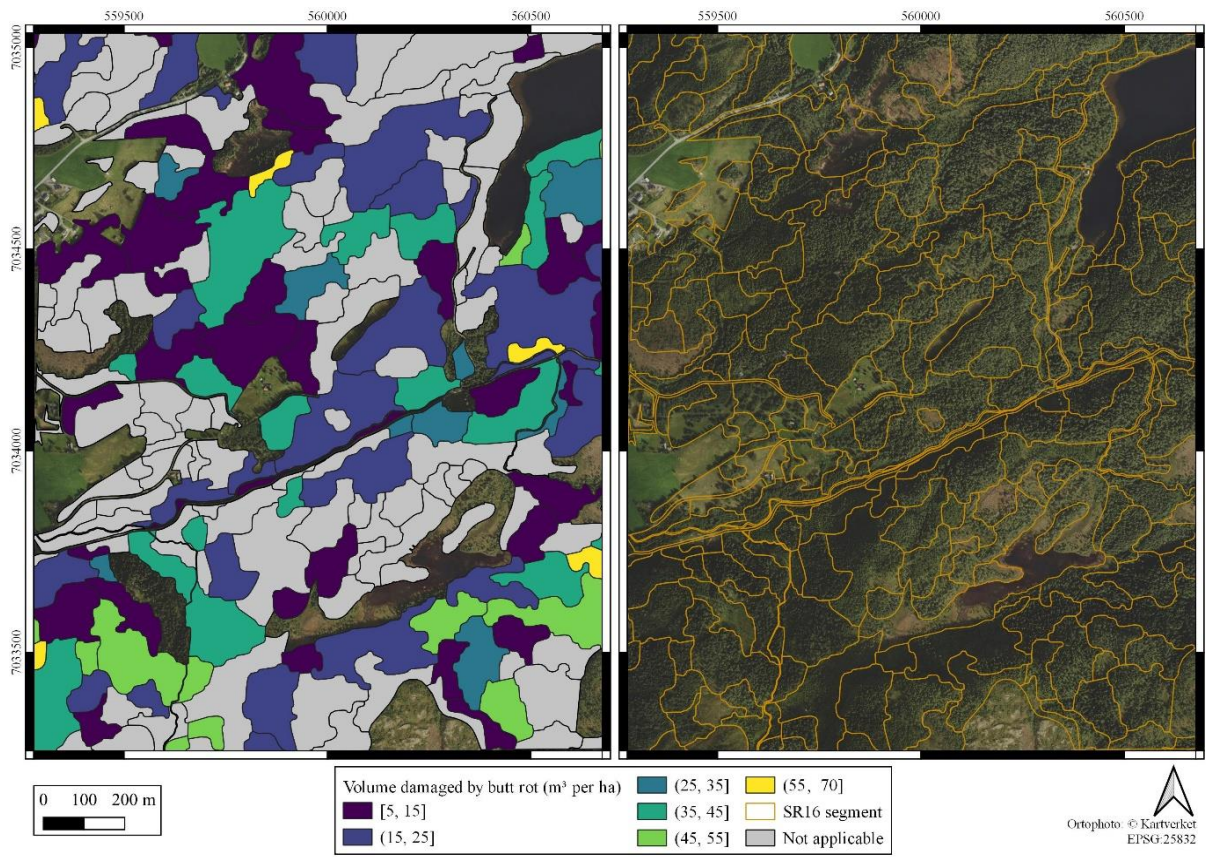

**Figure 4.** Mapping of predicted timber volume damaged by butt rot for stand segments of the Norwegian forest resource map SR16. "Not applicable" refers to non-mature forests or forests not dominated by spruce.

### 3.2 Importance of predictor variables

A list of the selected predictor variables and their importance in predicting butt rot is in Figure 5. The 95th percentile of the ALS height distribution ($H95_{ALS}$) was the most important variable. In the RF model with all predictor variables, harvested volume ($V_{HRV}$) was the second most important among all variables and the most important harvester variable. Harvester variables associated with the DBH distribution, especially quadratic mean DBH ($QMD_{HRV}$), were also observed to be important in the RF model. Harvested volume ($V_{HRV}$) and quadratic mean diameter ($QMD_{HRV}$) and the width of DBH distribution ($DR_{HRV}$) had Spearman correlations of larger than 0.5 with butt rot volume (Figure 6). Other variables with Spearman correlations of larger than 0.5 with the response were $H95_{ALS}$ and the variance of ALS heights ($Hvar_{ALS}$).

Remotely sensed variables were generally more important predictor variables than the environmental variables. Figure 5 and 6 show that the remotely sensed variables, especially ALS variables, are related to the response variable, and the variable importance values associated with remotely sensed variables were comparable with harvester variables. The most important environmental variable was the Y coordinate associated with the harvested stand. In terms of the Spearman correlation, slope ($SL_{TER}$) had the largest correlation among the environmental variables but its importance values associated with both RF models were small. The X and Y coordinates also had a weak correlation with the response indicating no strong spatial trend in butt rot abundance. The predictor variables were positively correlated with the response variable except the near-infrared band of Sentinel-2 ($NIR_{S2}$), distance to coast ($DC_{ENV}$), altitude ($AL_{ENV}$) and the Y coordinate of harvested stand (Y).

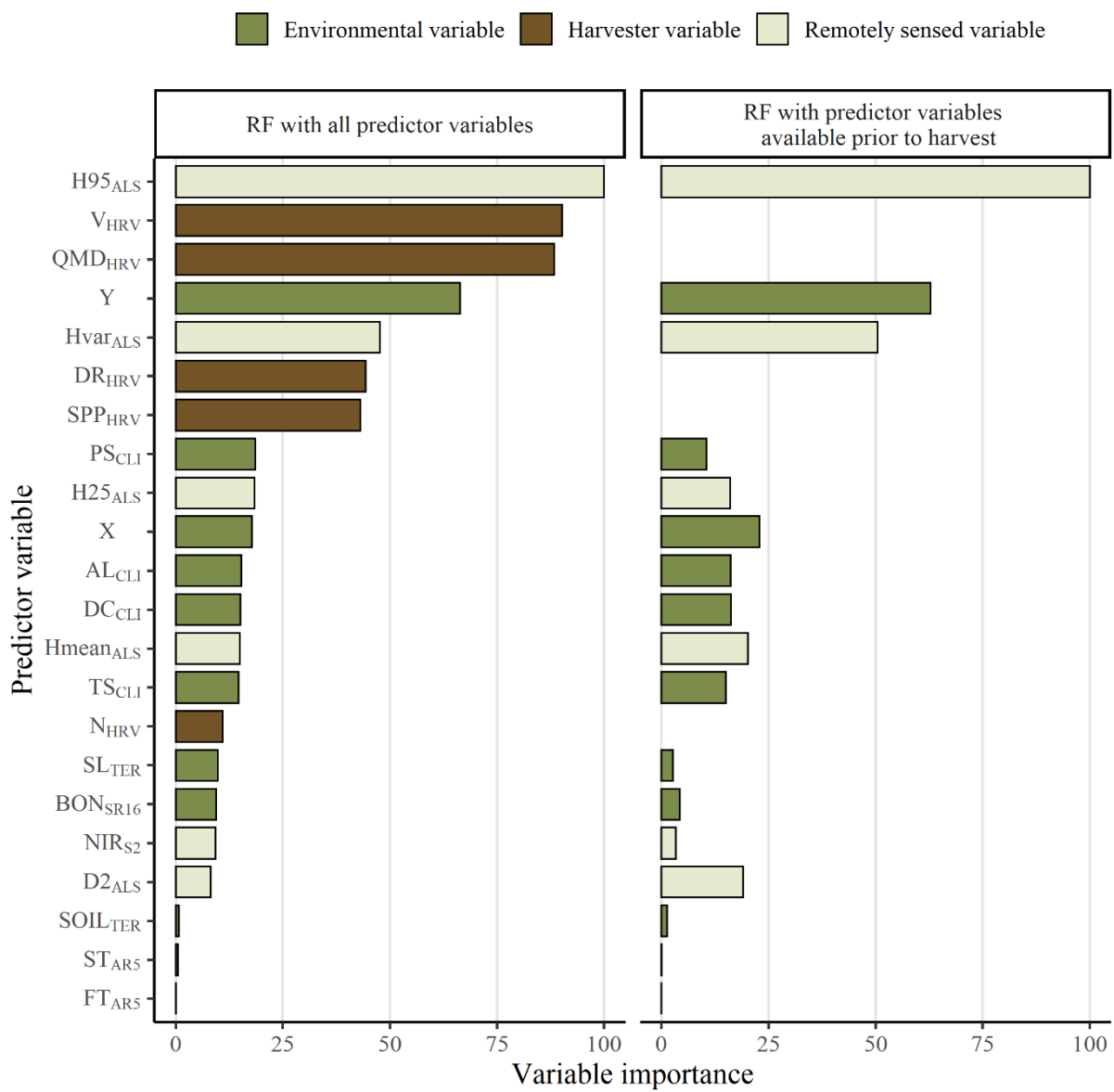

**Figure 5.** Variable importance values associated with predictor variables of the Random Forest (RF) models. See Table 2 for the description of the predictor variables.

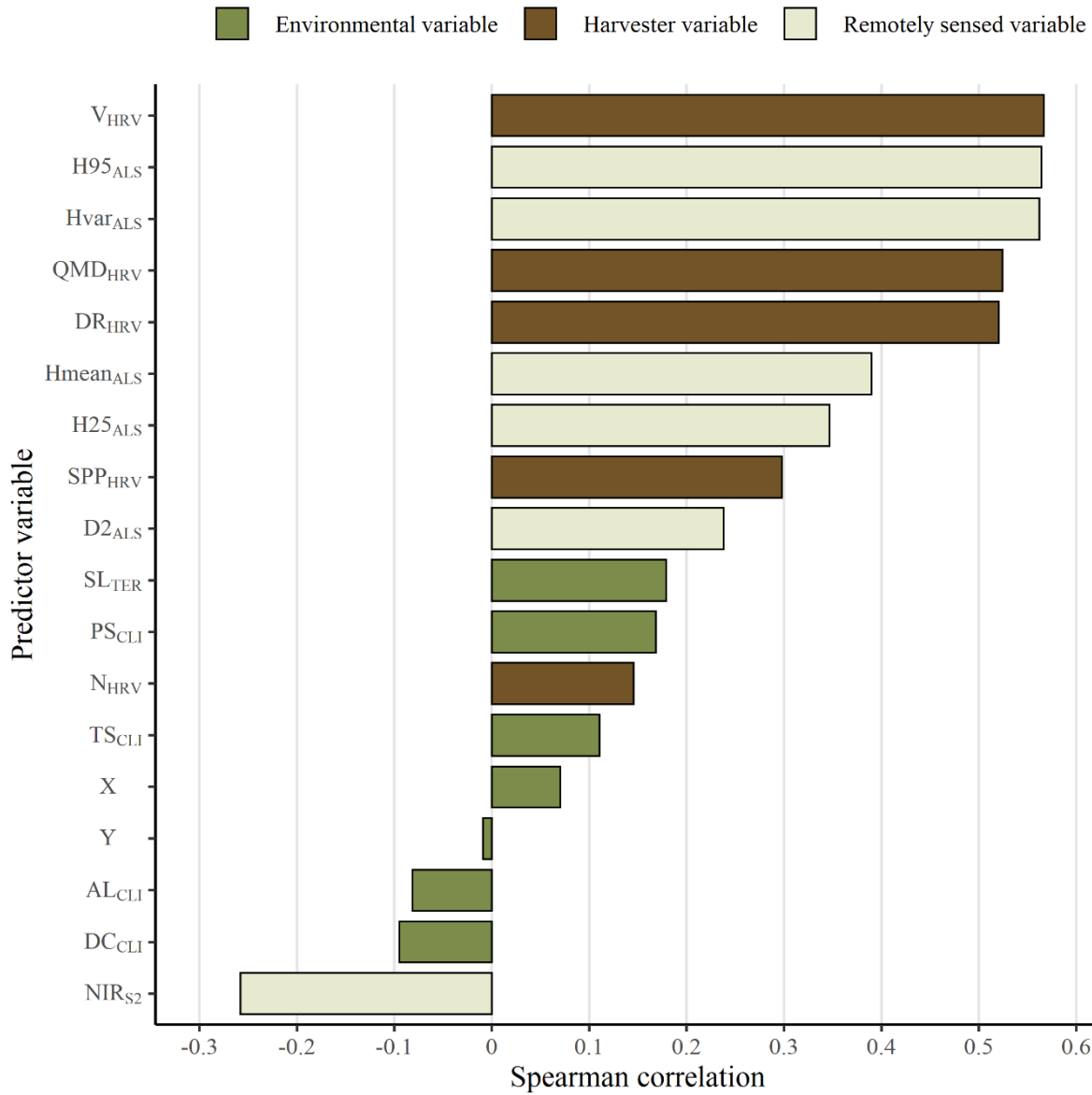

**Figure 6.** Spearman correlations between the response variable (volume damaged by butt rot m$^3$·ha$^{-1}$) and each numeric predictor variable used in the Random Forest regression. See Table 2 for the description of the predictor variables.

## 4 Discussion

We scrutinized the modeling and subsequent mapping of stand-level spruce BR volume using harvester data in Norwegian spruce-dominated forests. The recording of BR damages at the level of individual stems requires additional effort from the harvester operator since BR damages must be visually observed at the crosscuttings of stems and manually recorded. BR damages are therefore not routinely recorded during harvest operations in Norway so far. This study indicates that the harvester data are a potentially valuable source for the mapping of BR damages in mature spruce-dominated forests.

The harvester variables, namely volume and quadratic mean DBH, and remotely sensed variables extracted from ALS data ($H95_{ALS}$ and $Hvar_{ALS}$) were among the most important predictor variables. The abovementioned predictor variables are associated with the maturity of forest which is known to positively correlate with the risk of BR damages (Hylen and Granhus, 2018; Müller et al., 2018). The large importance of harvester variables indicates that there is potential to decrease error rates associated with BR volume predictions by improving forest attribute maps, such as timber volume (Rahlf et al., 2021) and DBH distributions (Räty et al., 2021).

We focused on Norway spruce-dominated forests which means that the harvested stands were not always pure Norway spruce stands. Mixed stands are linked to the slower spread of BR damages compared with monocultures (Möykkynen and Pukkala, 2010). It is also evident that the likelihood of observing a large absolute BR volume is higher when the spruce volume proportion is large. Therefore, it is important to employ predictor variables that provide information on the tree species composition in the model. The tree species composition can be mapped using remotely sensed variables, such as the optical bands of Sentinel-2, or indirectly, for example, with predictor variables characterizing growing conditions in forests (Breidenbach et al., 2020). Several predictor variables may indirectly provide information on tree species compositions, which may be the reason why the harvester data-based spruce volume proportion ($SPP_{HRV}$) was not among the most important predictor variables in this study. It should also be noted that the harvested stands used in this study were in average strongly dominated by Norway spruce (Table 1).

Hylen and Granhus (2018) found that BR damages were linked to temperature sum and altitude. In our study, the environmental variables were generally not as important predictor variables as harvester or remotely sensed variables. It is critical to note that our study had smaller geographical coverage and the forests were structurally more homogenous compared with the study of Hylen and Granhus (2018). These differences may lead to the underestimation of the predictive power associated with the environmental variables in this study. We also used the geographical coordinates of the harvested stands in the RF models and they achieved relatively large variable importance. However, it was found that the coordinates did not correlate with the response variable. Care must be taken when interpreting the variable importance values associated with the X and Y coordinates of the harvested stands. Their importance values do not directly indicate differences in terms of south-north or west-east directions, since the geographical coverage of our data was not comprehensive. The large importance values associated with the geographical coordinates rather resulted from the spatial autocorrelation associated with the BR observations among the harvested stands at the level of sub-regions.

StandCV resulted in smaller error rates than ClusterCV, which confirms our hypothesis regarding the spatial autocorrelation of the BR damages. This can be explained by the fact that a target forest and its geographically nearest harvested stands likely are similar in growing conditions and silvicultural history. Thus, the results suggest that geographically comprehensive harvester data are required in order to further reduce prediction errors of BR volume. Especially, care must be taken when creating BR volume maps for new forested areas without harvested reference stands nearby.

The harvester data were not a probability sample over the study area and have a selection bias towards clear-cut stands. This means that the harvester data are usually limited to mature forest stands, which affects the applicability of the models fit based on harvester data. Therefore, the use of harvester data is usually studied in the context of timber procurement which is associated with mature forests (Hauglin et al., 2018; Karjalainen et al., 2020; Peuhkurinen et al., 2008; Söderberg et al., 2021). There are also a few general challenges recognized in the application of harvester datasets. For example, the total timber volume is underestimated by harvesters (Kemmerer and Labelle, 2020), which is, however, not a problem in this study since the RB damage never reaches the top of a tree. Furthermore, retention trees are not typically recorded by the harvester, which may potentially decrease observed butt rot volumes at the stand-level. In addition, differences among the bucking schemes of the harvester machines (e.g. minimum allowed length of pulpwood log) may affect the accumulation of timber assortment volumes at the stand level among the operation areas. We did not have access to the bucking schemes used by the harvesters.

Due to the technological differences in the GNSS systems of the harvesters, the positioning errors vary among the harvesters. It is realistic to assume that the average positioning error of the harvester head-positioned trees in our dataset likely ranges between 5 m and 10 m. For the other trees without harvester head positions, the average positioning error is likely larger than 15 m. Harvested head-based locations were recorded for 52% of the trees which have the most accurate positions. An average positioning accuracy of 1 m can be achieved with an integrated positioning system which utilizes GNSS receivers and other sensors mounted in the harvester (Hauglin et al., 2017; Noordermeer et al., 2021). The positioning errors negatively affect the model errors of forest attributes and the effect increases with the decreasing size of the modeling units (Saukkola et al., 2019). We minimized the effect caused by positioning errors by using stands as the modeling units, removing small harvested stands and few recorded trees.

The findings of this study suggest that harvester data are a potential source for the mapping of BR volumes in mature spruce-dominated forests. We showed that a geographically comprehensive reference database is needed to minimize the error rates associated with the mapping of BR damages. Future work should consider different methodological solutions for the utilization of the continuous dataflow of harvester data in the mapping of BR damages.

# Conclusions

We draw the following conclusions from this study: (1) Volume damaged by butt rot can be mapped using observations from cut-to-length harvester data combined with remotely sensed and environmental predictor variables. (2) Geographically comprehensive training data from an area of interest are required to map butt rot damages with satisfactory accuracy. (3) Predictor variables that characterize the maturity of a forest stand, such as remote sensing-based height characteristics, were the most important predictor variables in the modeling of butt rot volume. (4) The use of forest attributes obtained from harvester data as predictor variables, in addition to the remotely sensed and environmental variables, decreased error rates, which suggests improved forest attribute maps may improve butt rot volume maps.

**Acknowledgements**

We express our gratitude to Simon Berg for the processing of harvester data. We would also like to thank Johannes Rahlf and Johannes Schumacher for their efforts in various data processing steps. We appreciate Ari Hietala's comments on the manuscript.

**Funding**

This study was supported by the Norwegian research council through the PRECISION project (NFR# 11067).